\documentclass[prb,titlepage,twocolumn]{revtex4}
\usepackage{amsmath}
\usepackage{amssymb}
\usepackage{bm}
\usepackage{graphicx}

\begin{document}

\title{Phase diagram for the transition from photonic crystals to dielectric metamaterials}

\author{Mikhail~V.~Rybin${}^{1,2}$}
\email{m.rybin@mail.ioffe.ru}
\author{Dmitry~S.~Filonov${}^{2}$}
\author{Kirill~B.~Samusev${}^{1,2}$}
\author{Pavel~A.~Belov${}^{2}$}
\author{Yuri~S.~Kivshar${}^{2,3}$}
\author{Mikhail~F.~Limonov${}^{1,2}$}

\affiliation{$^1$Ioffe Physical-Technical Institute, St.~Petersburg 194021, Russia\\
$^2$ Department of Nanophotonics and Metamaterials, ITMO University, St.~Petersburg 197101, Russia\\
$^3$Nonlinear Physics Center, Australian National University, Canberra ACT 0200, Australia }

\begin{abstract}
Photonic crystals and metamaterials represent two seemingly different classes of artificial electromagnetic media but often they are composed of similar structural elements arranged in periodic lattices. The important question is how to distinguish these two types of periodic photonic structures when their parameters, such as dielectric permittivity and lattice spacing, vary continuously. Here, we discuss transitions between photonic crystals and all-dielectric metamaterials and introduce the concept of a phase diagram and an order parameter for such structured materials, based on the physics of Mie and Bragg resonances. We show that a periodic photonic structure transforms into a metamaterial when the Mie gap opens up below the lowest Bragg bandgap where the homogenization approach can be justified and the effective permeability becomes negative. Our theoretical approach is confirmed by detailed microwave experiments for a metacrystal composed of a square lattice of glass tubes filled with heated water. This analysis yields deep insight into the properties of periodic photonic structures, and it also provides a useful tool for designing different classes of electromagnetic materials in a broad range of parameters. 
\end{abstract}

\date{\today}


\maketitle



The concepts of phase transitions and phase diagrams are among the most fundamental concepts in physics. Here we reveal and analyse a rather unusual type of phase transitions between two classes of artificial electromagnetic structures: photonic crystals (PhCs) \cite{g101,g142} and metamaterials (MMs) \cite{enghata2006electromagnetic,smith2010metamaterials}. Rapidly growing attention to the study of PhCs and MMs is due to their unique electromagnetic properties and the exciting new physics these structures may offer for novel applications in photonics \cite{cai2007optical,lapine2012magnetoelastic,silva2014performing,smith2004metamaterials,soukoulis2006negative}. The resonant properties of PhCs are driven by the familiar Bragg scattering from a periodic structure when the lattice spacing becomes comparable with the wavelength of light. MMs are regarded as effective media with averaged parameters which in turn can be engineered by changing the properties of their subwavelength elements modifying the propagation of electromagnetic waves with the wavelengths much larger than the lattice spacing.

Many properties of conventional MMs are usually defined by the optically-induced magnetic response of metallic wires or split-ring resonators, which has been demonstrated experimentally in many studies for both microwave and near-infrared spectral regions \cite{soukoulis2011past,zheludev2012metamaterials}. However, the scaling of such MMs to higher frequencies encounters many problems, including strong absorption of electromagnetic radiation in metals. Recently suggested electromagnetic periodic structures composed of high-permittivity dielectric elements \cite{huang2004negative,kallos2012resonance,moitra2013realization} can avoid these problems, while still delivering an effective magnetic response. The main physics underlying the properties of such all-dielectric MMs is based on the resonant modes of the Mie scattering \cite{g120} supported by dielectric particles \cite{ginn2012realizing,zhao2009mie}. Mie resonances provide a mechanism for the generation of an optically-induced resonant magnetic response based on the displacement current, and they offer a simpler and more versatile route towards the fabrication of isotropic MMs and metasurfaces operating at optical frequencies. 

PhCs and all-dielectric MMs can be composed of the similar structural elements arranged in the similar geometries. Such structures offer a unique opportunity to study a transition from PhCs to MMs when their parameters, such as dielectric permittivity and lattice period, vary continuously. The transition we describe here can be mapped onto a phase diagram; however it is quite different from phase transitions well known in other fields such as thermodynamics. The thermodynamical phase transitions appear as transformations of a system from one state or ``phase'' to another with changing external variables such as pressure, temperature, chemical potential, etc. Here we reveal and analyze a rather unusual type of phase transitions between two regimes of light propagation, i.e. the regime where the propagation of light is mainly influenced by the interference between multiple Bragg scattered (PhC ``phase''), and the regime where the propagation of light is mainly determined by the properties of each single element of the materials (MM ``phase''). 

\begin{figure}[!t]
\includegraphics{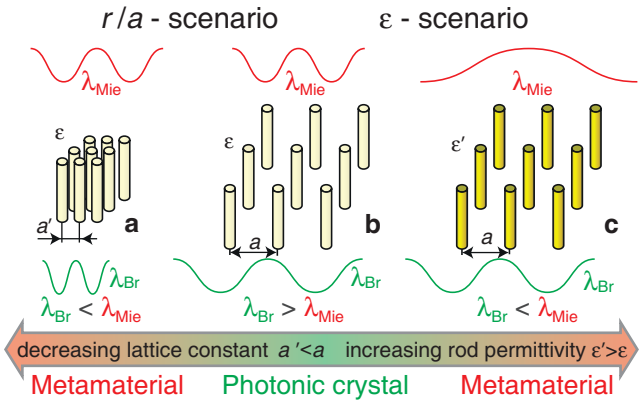}
\caption{
\textbf{Transition from photonic crystals to metamaterials.} Schematic illustrates two possible transformations of the PhC \textbf{b} to MM \textbf{a},\textbf{c} through the competition between the Bragg and Mie scattering wavelengths. To satisfy the classical homogenization condition in the low-frequency long-wavelength limit ($\lambda _{Bragg} <\lambda _{Mie} $), one may decrease the lattice spacing and increase the filling ratio $r/a$ ($r/a$-scenario) or, alternatively, increase the rod permittivity $\varepsilon $ ($\varepsilon $-scenario).
}
\label{fig:scheme}
\end{figure}

First, we identify two different scenarios for the transition of a two-dimensional square lattice of dielectric rods from PhC to MM (see Fig. 1). The first scenario that has never been discussed previously occurs when the lattice spacing $a$ decreases in comparison with the fixed radius of rods $r$ and dielectric permittivity $\varepsilon $. This transformation leads to an increase of the filling ratio $r/a$, and it makes possible the homogenization \cite{andryieuski2010homogenization} of the periodic dielectric structure with the negative effective permeability ($\mu <0$) at higher values of $\varepsilon $ \cite{o2002photonic,bron2007,VynckPRL2009}. 

The second scenario occurs when the dielectric permittivity of rods $\varepsilon $ varies in a fixed square lattice. This transformation leads to an increase of the Mie scattering wavelength, and it can also provide the conditions for the homogenization approach with negative $\mu $. While photonic structures with varying $\varepsilon $ were studied earlier theoretically \cite{fote2012,notomi2000theory}, the transition between PhCs and MMs has never been analyzed previously to our best knowledge, especially in terms of the phase diagram, as discussed below.

\section*{RESULTS}
\subsection*{The interplay between Bragg and Mie resonances}

To analyze the PhC -- MM phase transition, we investigate theoretically electromagnetic properties of the dielectric periodic medium depending on the dielectric contrast, filling ratio and incident wave polarization. As an illustrative example, we consider here 2D square lattices of parallel dielectric circular rods infinitely long in the $z$ direction with the radius $r$ and real frequency independent permittivity $\varepsilon $. The rods are embedded in air ($\varepsilon _{air}=1$) and the arrays can be characterized by the filling ratio $r/a$. We have derived three key sets of spectroscopic data (see Methods), namely: (i) spectra of the Mie scattering by an isolated rod; (ii) photonic band structure of an infinite 2D square lattice composed of rods; and (iii) transmittance of a 2D square lattice of 10 layers in length. All data sets have been calculated for a wide range of the rod permittivity $1{\rm \leqslant }\varepsilon {\rm \leqslant }100$ with the step of $\Delta \varepsilon =1$ ($\varepsilon $-scenario) and in the range of filling ratio $0{\rm \leqslant }r/a{\rm \leqslant }0.5$ with the step of 0.01 ($r/a$-scenario). Note that at $r/a=0.5$ all rods touch each other and at $r/a>0.5$ the rods penetrate into each other and the structure appears as inverted one of air holes in a dielectric matrix. The calculated data (see Methods) allow us to analyze precisely the evolution of the low-frequency region of the photonic band structure formed by a complicated mixture of the dispersion curves originating from Mie and Bragg resonances and propagating modes. Since our goal is to discuss the PhC -- MM transition, we concentrate on the TE polarization where the lowest TE$_{01} $ magnetic dipole resonance of the rods results in a negative magnetic permeability in the Mie gap spectral region \cite{o2002photonic}.

The physics of the PhC -- MM transition is displayed by the gap map (Fig. 2) that relates the Bragg and Mie gaps positions to the rod permittivity $\varepsilon $ for given $r/a=0.25$. These illustrate essential features such as strong decreasing of Mie resonance frequencies with $\varepsilon $ increase followed by intersection of the Mie and Bragg bands. The Mie resonances of individual rods form both non-dispersive flat bands in the photonic structure and Mie photonic gaps in the energy spectrum (Figs. 2,6). For low-contrast PhCs, all Mie resonances are located is a frequency range higher than the lower Bragg gap. With increasing rod permittivity $\varepsilon $, the Mie resonances of single rods and correspondingly Mie bands of the structure demonstrate a shift from the higher to low frequencies, crossing the strongly dispersive Bragg bands. The creation of the lowest Mie gap driving the artificial magnetism is clearly seen both in the band structure and in the transmission spectra (Figs. 2,6).

\begin{figure*}[!t]
\includegraphics{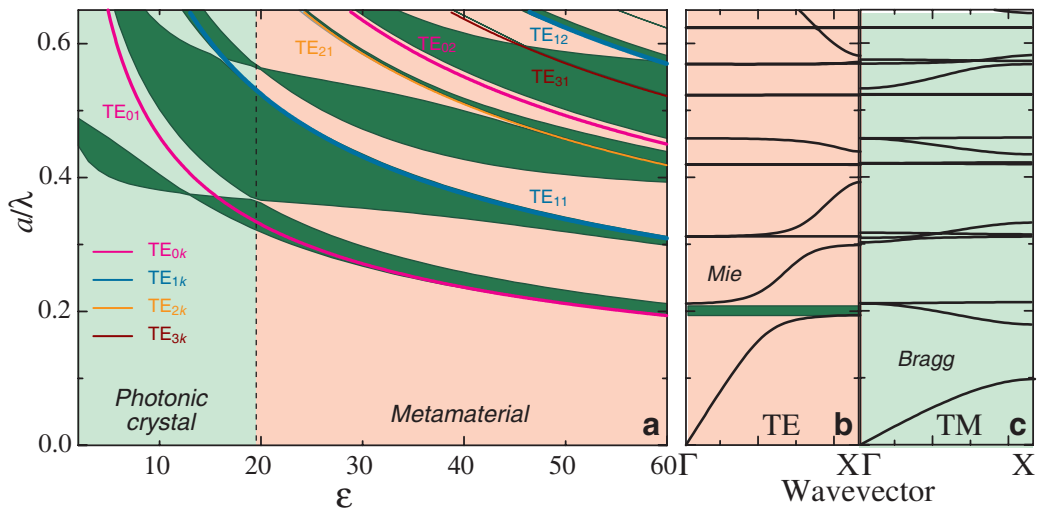}
\caption{
\textbf{Bandgap diagram for the 2D square lattice of circular rods.} \textbf{a} The bandgap diagram obtained from the photonic band structures calculated for $1{\rm \leqslant }\varepsilon {\rm \leqslant }60$ with steps of $\Delta \varepsilon =1$. The rods are embedded in air, $r/a=0.25$, and TE polarization is considered. The Bragg- and Mie-gaps obtained from the band structure calculations are marked by dark green. (\textbf{b}, \textbf{c}) The photonic band structure for the 2D square lattice of rods with $\varepsilon =60$ for TE- and TM-polarization respectively for the $\Gamma \to X$ scan of the wave vector $k$.
}
\label{fig:GapMap}
\end{figure*}

\begin{figure}[!t]
\includegraphics{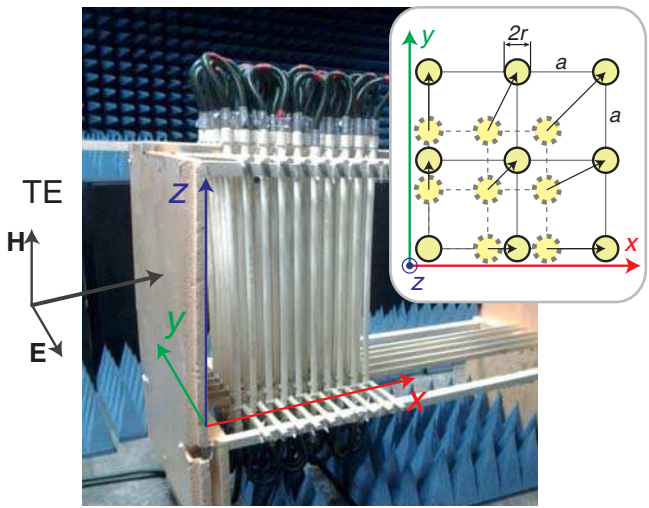}
\caption{
\textbf{Experimental setup for transmission measurements and schematic of the lattice transformation.} An anechoic chamber with a metacrystal composed of 10 tubes in length ($x$-axis) and 5 tubes in width ($y$-axis), 50 tubes in total. Each tube has a length of 1 m (along $z$-axis), an outer radius of 2 cm and an inner radius of 1.7 cm. All tubes are connected in a closed series circuit together with a water heater and a temperature stabilizer. Insert: Schematic illustration of the transformation of the 2D square lattice composed of dielectric rods as the lattice constant $a$ increase. The scheme shows the square lattice from above. The rods with dielectric permittivity $\varepsilon $ and fixed radius $r$ are embedded in air ($\varepsilon _{air} =1$). The material is homogeneous along the $z$ direction and periodic along $x$ and $y$. 
}
\label{fig:Setup}
\end{figure}

\begin{figure*}[!t]
\includegraphics{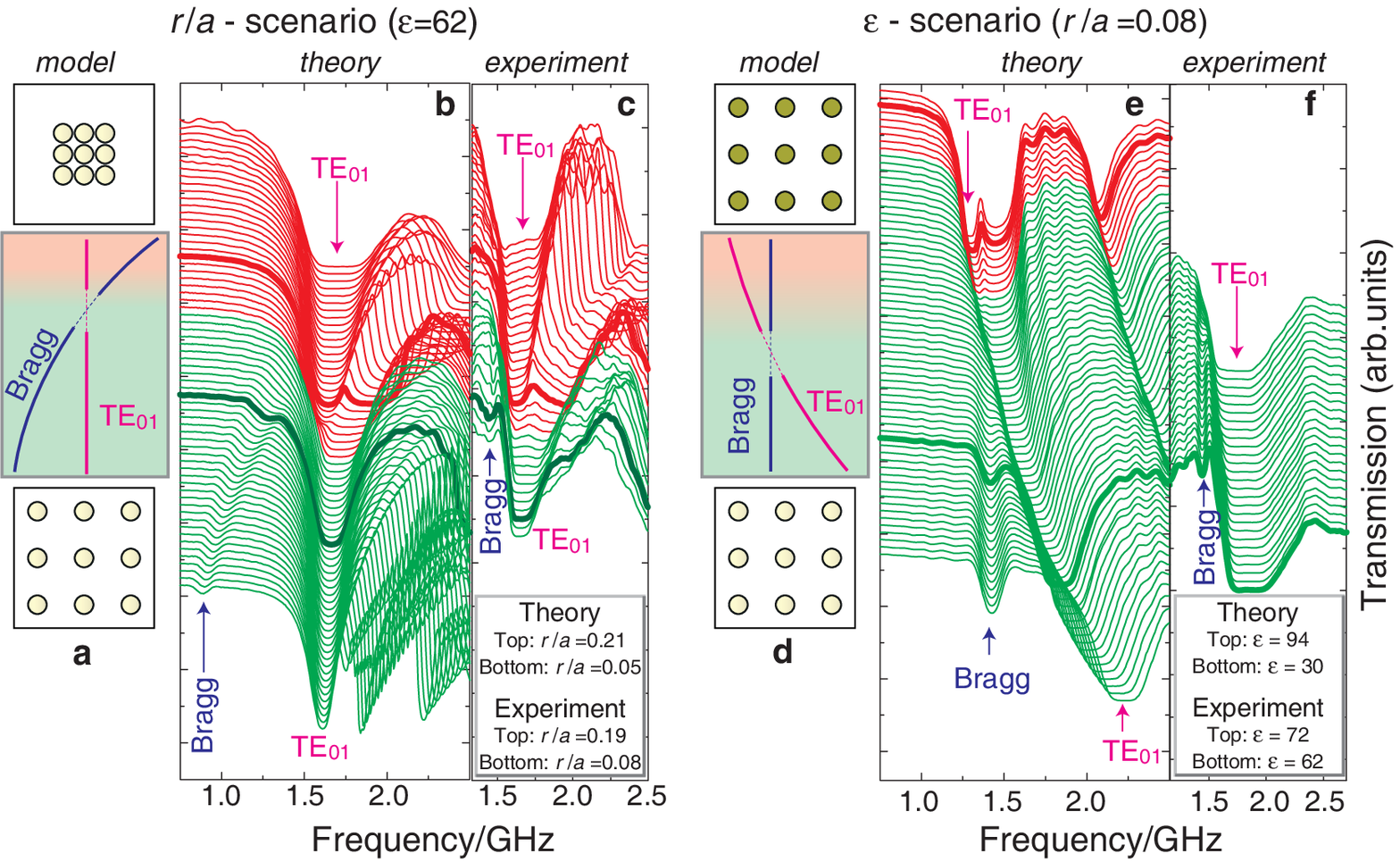}
\caption{
\textbf{Transmission spectra for a square lattice of circular rods: Experiment and numerics.} TE polarization. \textbf{a}-\textbf{c} $r/a$-scenario, $\varepsilon =62$. The calculated \textbf{b} and experimentally measured \textbf{c} transmission spectra of the metacrystal as a function of the filling ratio $r/a$. The calculated spectra are shown in the range $0.05{\rm \leqslant }r/a{\rm \leqslant }0.21$, the experimental spectra are shown in the range $0.08{\rm \leqslant }r/a{\rm \leqslant }0.19$. \textbf{d}-\textbf{f} $\varepsilon $-scenario, $r/a=0.08$. The calculated \textbf{e} and experimentally measured \textbf{f} transmission spectra of the metacrystal as a function of the dielectric permittivity. The calculated spectra are shown in the range $30{\rm \leqslant }\varepsilon {\rm \leqslant }94$, the experimental spectra are shown in the range $62{\rm \leqslant }\varepsilon {\rm \leqslant }94$. The spectra are shifted vertically by the constant value. 
}
\label{fig:Spectra}
\end{figure*}

\subsection*{A metacrystal: two crossover scenarios}

Armed with the results of our calculations, we verify our concept experimentally by engineering a metacrystal composed of plastic circular tubes filled with water and forming a 2D square lattice with variable lattice constant $a$ (Fig. 3). The structure of the metacrystal enables us to perform two types of experiments according to the $r/a$ - and $\varepsilon $-scenarios. The $r/a$-scenario was realized using a special construction of the metacrystal that makes it possible to change the distance between tubes synchronously in $x$ and $y$ directions keeping the 2D square symmetry of the lattice (Fig. 3). For the $\varepsilon $-scenario, we employ the advantages of strong temperature dependence of the dielectric permittivity of water that is characterized by $\varepsilon =80$ at 20$^{\circ } $C and $\varepsilon =50$ at 90$^{\circ } $C in the microwave frequencies range from 1 GHz to 6 GHz \cite{kaatze1989complex}. Based on the calculated results and the parameters of the metacrystal, we chose intervals of most interest for each scenario, namely $60{\rm \leqslant }\varepsilon {\rm \leqslant }80$ at $0.7{\rm \leqslant }r/a{\rm \leqslant }0.8$ for the $\varepsilon $-scenario and $0.65{\rm \leqslant }r/a{\rm \leqslant }1.9$ at $\varepsilon =60$ for the $r/a$-scenario.

The transmission spectra of the metacrystal installed in an anechoic chamber were measured in the 1 GHz to 3 GHz range. A rectangular horn antenna (TRIM 0.75 GHz to 18 GHz; DR) connected to a transmitting port of the vector network analyzer Agilent E8362C was used to approximate plane-wave excitation. A similar horn antenna was employed as a receiver. For the electromagnetic wave incident perpendicular to the tube axis $z$, the two transverse polarizations appear decoupled and the vector wave problem can be reduced to two independent scalar equations. As a result, the modes of 2D PhCs can be classified as either transverse-electric modes TE ($E_{x} $, $E_{y} $, $H_{z} $) with the electric field confined to the $x-y$ plane and the magnetic field polarized along the axes of the rods, or transverse-magnetic modes TM ($H_{x} $, $H_{y} $, $E_{z} $) for which the magnetic field is oriented perpendicular to the rod axis $z$. For the TM polarization such structures can not show any magnetic activity \cite{Ramakrishna2005}, so we only consider TE-polarization.

Figure 4 demonstrates excellent agreement between the measured and calculated transmission spectra both for $r/a$ - and $\varepsilon $ - dependencies. For the $r/a$-scenario, we started our experiment from the PhC phase ($r/a=0.08$) and compressed the structure by decreasing the lattice constant $a$ (Fig. 4\textbf{a}). The lowest Bragg gap demonstrates remarkable shift towards high-frequency region, then nearly almost disappears at $r/a\approx 0.09$ when it approaches the unchanging TE$_{01} $ Mie gap from the low-frequency side, and finally appears again at $r/a\approx 0.095$ on the high-frequency side of the Mie gap. As a result, the lowest Bragg band and the TE$_{01} $ Mie band change their positions in the energy scale inaugurating the transition into MM phase. In contrast, for the $\varepsilon $-scenario, the Bragg gap position changes very slowly as opposed to the strong shift of the TE$_{01} $ Mie band that finally leads to the same general result (Fig. 4\textbf{d}).

\begin{figure*}[!t]
\includegraphics{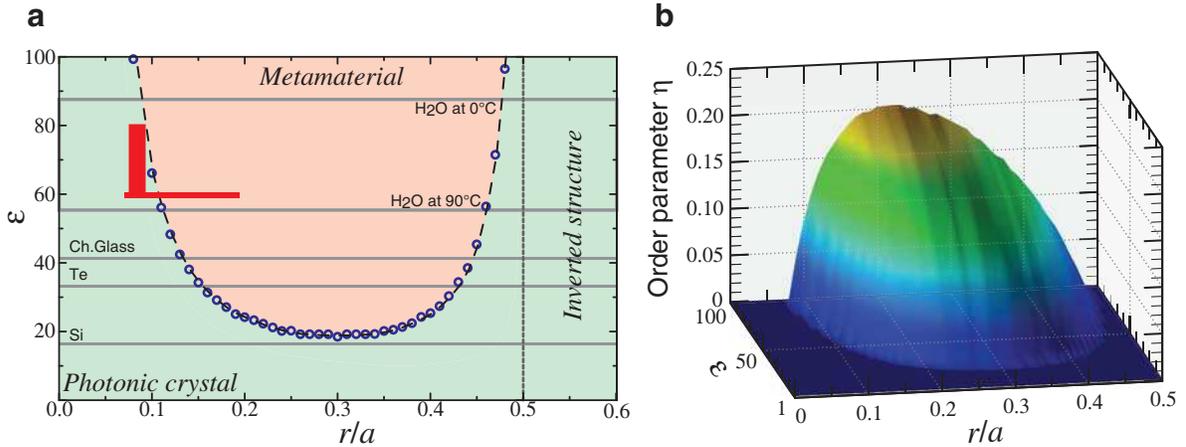}
\caption{
\textbf{PhC-MM phase diagram.} \textbf{a} PhC-MM phase diagram for a square lattice of dielectric circular rods. TE polarization. Blue circles indicate the points where the TE$_{01} $ Mie gap splits from Bragg band becoming the lowest gap in the spectrum (Fig. 6). The experimentally investigated regions are marked by red. The dielectric permittivities $\varepsilon $ of different materials are marked by horizontal lines. \textbf{b} The dependence of the order parameter $\eta $ on the dielectric permittivity $\varepsilon $ and the filling factor $r/a$. The order parameter is defined as a difference in the energy between the TE${}_{01}$ Mie and Bragg gaps in the MM phase.
}
\label{fig:PhaseDiag}
\end{figure*}

\section*{DISCUSSION}

Figure 5\textbf{a} shows our main result: the PhC-MM phase diagram for the 2D square lattice of circular rods. The axes represent the dielectric permittivity $\varepsilon $ and filling ratio $r/a$. The phase diagram was obtained based on 50 gap maps (such as in Fig. 2) calculated in the permittivity region $1{\rm \leqslant }\varepsilon {\rm \leqslant }100$ with the step $r/a=0.01$ for TE polarization. The diagram is divided into two areas, which represent the PhC and MM phases. The transition from PhC-phase to MM-phase can be identified by critical points which are defined by a special combination of permittivity $\varepsilon $ and filling ratio $r/a$. The curvature of the second dispersion curve around the $\Gamma $ point decreases and finally becomes perfectly flat along the $\Gamma $X path (at $\varepsilon =19.5$, $r/a=0.25$, Fig. 6\textbf{d}). With future increase in $\varepsilon $ or change in $r/a$, the TE$_{01} $ Mie gap completely splits from complicated coupled Mie-Bragg band becoming the lowest gap in the spectrum and the MM-phase appears. Within this Mie gap the magnetic permeability is negative and all waves in the medium are evanescent. Since all the Bragg gaps are located at higher frequencies, any diffraction losses in the vicinity of the Mie resonance are absent. The transition from the PhC phase to the MM phase is completed and the medium is classified as the MM with $\mu <0$. A phase transition is usually characterized by an order parameter that basically arises from symmetry breaking but can also be defined for non-symmetry-breaking transitions \cite{gooding1991role}. Here for the unusual phase transition PhC$\to$MM we define the order parameter $\eta $ as a measure of splitting in the energy scale between the TE${}_{01}$ Mie and Bragg gaps. It is zero in the PhC phase and nonzero in the MM phase. Figure 5\textbf{b} presents the order parameter depending on both external variables $\varepsilon $ and $r/a$.

Now we review some features of the PhC-MM phase diagram. At low filling ratio $r/a<0.05$, the concentration of the completely isolated rods in the lattice is too small for achieving negative average permeability at reasonable values of $\varepsilon $. In contrast, when distance between rods become very small ($r/a>0.45$), the inter-rod interaction become strong and the hopping of the Mie photons from one rod to another dominates. This means that Mie resonances are not localized anymore and they finally exhibit the character of propagating Bloch waves. All requirements for the appearance of the MM-phase are broken and the PhC-phase extends its region of existence to very high $\varepsilon $.

Identification of candidate dielectric low-loss materials is critical in designing a practical MM for different ranges of electromagnetic spectrum. As an example, we plot in Fig. 5\textbf{a} the regions of the MM-phase for water in microwave range (region $55{\rm \leqslant }\varepsilon {\rm \leqslant }87$) \cite{Ramakrishna2005}, chalcogenide glasses in IR ($\varepsilon \approx 41$) \cite{nvemec2009ge} and tellurium in visible ($\varepsilon \approx 33$). The 2D square lattice of silicon ($\varepsilon \approx 16$) \cite{palik1998handbook} rods has only PhC-phase.

\section*{Conclusions}

In this paper we integrate a 2D square structure composed of dielectric circular rods within a general phase diagram in the permittivity-filling ratio plane. The boundaries of the MM-phase were obtained theoretically and confirmed experimentally. Based on the proposed approach one can obtain different PhC-MM phase diagrams by altering the dimensions, symmetry, composition, size and geometry of structural elements within a cell. The underlying mechanism of the PhC-MM transition provide unique criterion for goal-oriented search of novel low - loss MM with a wide range of applications at infrared and visible frequencies.

{\em Acknowledgements.} We acknowledge fruitful discussions with W. Barnes, I. Brener, S. Fan, S. Foteinopoulou, A. Khanikaev, D.A. Powell, J.R. Sambles, and A.P. Slobozhanyuk and thank S.B. Glybovski for technical help. This work was supported by the Government of the Russian Federation (grant 074-U01), Russian Foundation for Basic Research (grant 15-02-07529), Dynasty Foundation, and the Australian Research Council.


\section*{Author contributions}

M.R. and K.S. developed the theoretical model and conducted simulations and data analysis. D.F. performed experimental measurement. M.L, P.B. and Y.K. provided guidance on the theory, numerical analysis and experiment. All authors discussed the results and contributed to the writing of the manuscript.

\section*{Additional information}
The authors declare no competing financial interests.
Correspondence and requests for materials should be addressed to M.R. (Email: m.rybin@mail.ioffe.ru)

\section*{METHODS}


\begin{figure*}[!t]
\includegraphics{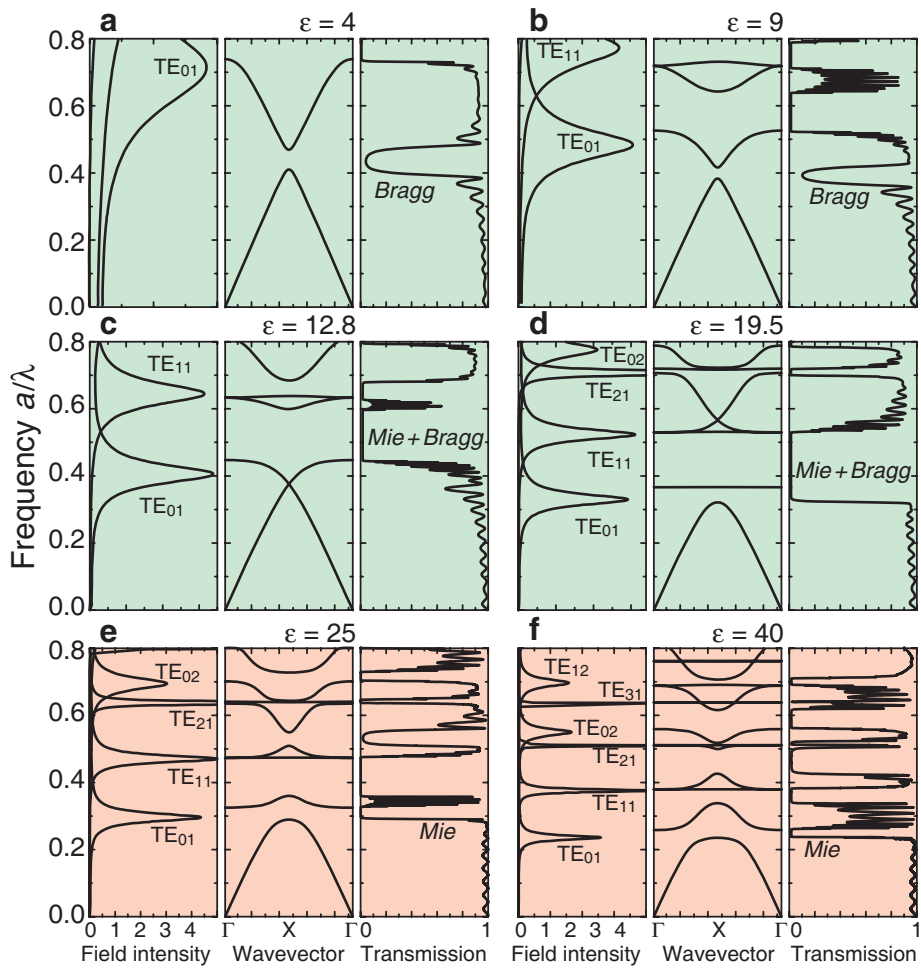}
\caption{
(Left-hand sections in all panels) Calculated Mie scattering efficiency $Q_{sca,n} $ for an isolated dielectric circular rod for TE$_{n} $ (${\rm n}\ge 0$) modes. (Central sections) The band structure for 2D square lattice of rods with $r=0.25a$ in air ($\varepsilon _{air}=1$) for the TE polarization. The band structure is shown between the $\Gamma $ point (wave vector $k=0$) and the X point ($\left|k\right|=\pi /a$ along the $x$ direction). (Right-hand panels) The transmittance calculated for 10 lattice layers of the 2D square structure of rods in air for the TE polarization. The frequency and wave vector are plotted in dimensionless units $a/\lambda $, where $\lambda $ is vacuum wavelength and $a$ denotes the lattice constant. \textbf{a} $\varepsilon =4$, \textbf{b} $\varepsilon =9$, \textbf{c} $\varepsilon =12.8$, \textbf{d} $\varepsilon =19.5$, \textbf{e} $\varepsilon =25$, \textbf{f} $\varepsilon =40$. 
}
\label{fig:FullPicture}
\end{figure*}

The scattering properties of an isolated dielectric circular rod and 2D square lattices composed from such rods were calculated using three independent methods. We calculated (i) spectra of the Mie scattering by an isolated rod; (ii) photonic band structure of a 2D square lattice composed of rods; and (iii) transmittance of a 2D square lattice composed of 10 rods in length (along $x$-axis). All data sets calculated for a wide range of the rod permittivity $1{\rm \leqslant }\varepsilon {\rm \leqslant }100$ with the step of $\Delta \varepsilon =1$ and in the range of filling ratio $0{\rm \leqslant }r/a{\rm \leqslant }0.5$ with the step of $0.01$.

First we consider the Mie scattering by an isolated rod. The far-field scattering can be expanded into orthogonal electromagnetic dipolar and multipolar modes, described by circular Lorenz-Mie resonant coefficients $a_{n} $ (TE polarization) and $b_{n} $ (TM polarization) corresponding to electric and magnetic moments, respectively \cite{g120}. Figure 6 presents the numerically calculated spectra of the Mie scattering efficiency $Q_{sca,n} =\frac{2}{x} \left|a_{n} \right|^{2} $ for the dipole TE$_{0} $ and multipole TE$_{n} $ ($n{\rm \geqslant }1$) modes in the range of dimensionless frequencies $a/\lambda $ from 0 to 0.8. Each mode consists of an infinite set of quasi-equidistant harmonics TE$_{nk} $ ($k{\rm \geqslant }1$). The calculations reveal a strong decrease of the Mie eigenfrequencies and narrowing of the resonant Mie bands with $\varepsilon $ increasing, as can be clearly seen for the lowest TE$_{01} $ magnetic Mie resonance in Fig. 6.

The photonic band structures for the 2D square lattice of circular rods were computed numerically using the plane wave expansion method \cite{g405}. The square 2D lattice has a square Brillouin zone, with the three special points $\Gamma $, X, and M corresponding to $k=0$, $k=\frac{\pi }{a} \hat{x}$, and $k=\frac{\pi }{a} \hat{x}+\frac{\pi }{a} \hat{y}$, respectively. We restrict here our discussion to the case for which the wave vector $k$ of the eigenmodes is oriented in the $\Gamma \to $X direction of the first Brillouin zone displaying the lowest Bragg frequencies in comparison with the $\Gamma \to $M direction (Fig. 6). The photonic band structures were obtained with 128 by 128 plane waves.

The transmission spectra were calculated by using the CST Microwave Studio software for the wave vector of the incident beam parallel to the $\Gamma \to $X direction. We considered the 2D square lattice with different thicknesses along $x$ axis which leads to creation of the photonic band gaps, including thicknesses of 10 lattice layers as shown in Fig. 6.


\end{document}